\documentclass[12pt]{article}  
\usepackage{epsfig}
\usepackage{xspace}
\usepackage{pennames}
\usepackage{rotating}
\usepackage{cite}
\newcounter{enumct}
\newenvironment{Enumerate}{\begin{list}{\arabic{enumct}.}%
{\usecounter{enumct}\setlength{\topsep}{0.2mm}%
\setlength{\partopsep}{0.2mm}\setlength{\itemsep}{0.2mm}%
\setlength{\parsep}{0.2mm}}}{\end{list}}
\newcommand{\Z      }[4]{\ensuremath{#1\,\pm #2\,^{+\,#3}_{-\,#4}}\xspace}
\newcommand{\ZZ     }[2]{\ensuremath{^{#1}_{#2}}\xspace}
\newcommand{\epem   }{\ensuremath{\mathrm{e}^+\mathrm{e}^-}\xspace}
\newcommand{\ee     }{\ensuremath{\mathrm{ee}}\xspace}
\newcommand{\aem    }{\ensuremath{\alpha}\xspace}
\newcommand{\aemsq  }{\ensuremath{\aem^2}\xspace}
\newcommand{\pz     }{\ensuremath{\phantom{-}}\xspace}
\newcommand{\invpb  }{\ensuremath{\mathrm{pb}^{-1}}\xspace}
\newcommand{\eb     }{\ensuremath{E_\mathrm{b}}\xspace}
\newcommand{\qsq    }{\ensuremath{Q^{2}}\xspace}
\newcommand{\wsq    }{\ensuremath{W^{2}}\xspace}
\newcommand{\psq    }{\ensuremath{P^{2}}\xspace}
\newcommand{\qzm    }{\ensuremath{\langle \qsq \rangle}\xspace}
\newcommand{\ftn    }{\ensuremath{F_{2}^{\gamma}/\aem}\xspace}
\newcommand{\ftqn   }{\ensuremath{\ft(\qsq)/\aem}\xspace}
\newcommand{\ft     }{\ensuremath{F_{2}^{\gamma}}\xspace}
\newcommand{\fl     }{\ensuremath{F_\mathrm{L}^{\gamma}}\xspace}

\newcommand{\ftxq   }{\ensuremath{\ft(x,\qsq)}\xspace}
\newcommand{\flxq   }{\ensuremath{\fl(x,\qsq)}\xspace}
\newcommand{\ftxqp  }{\ensuremath{\ft(x,\qsq,\psq)}\xspace}
\newcommand{\der    }{\ensuremath{{\mathrm d}}\xspace}
\newcommand{\kt     }{\ensuremath{k_{\mathrm{t}}}\xspace}
\newcommand{\pt     }{\ensuremath{p_{\mathrm{t}}}\xspace}

\newcommand{\ea     }{\ensuremath{E_\mathrm{a}}\xspace}
\newcommand{\nch    }{\ensuremath{N_{\mathrm{trk}}}\xspace}

\newcommand{\etag   }{\ensuremath{E_{\mathrm{tag}}}\xspace}
\newcommand{\ttag   }{\ensuremath{\theta_{\rm tag}}\xspace}
\newcommand{\gev    }{\ensuremath{\mathrm{GeV}}\xspace}
\newcommand{\gevsq  }{\ensuremath{\mathrm{GeV}^2}\xspace}

\newcommand{\ssee   }{\ensuremath{\sqrt{s_{\rm e e}}}\xspace}

\newcommand{\GG     }{\ensuremath{{\gamma\gamma}}\xspace}
\newcommand{\Wvis   }{\ensuremath{W_{\mathrm{vis}}}\xspace}
\newcommand{\xvis   }{\ensuremath{x_{\mathrm{vis}}}\xspace}
\newcommand{\zn     }{\ensuremath{\mathrm{Z}^0}\xspace}
\newcommand{\znhad  }{\ensuremath{\zn/\gs\to\mbox{hadrons}}\xspace}
\newcommand{\zntau  }{\ensuremath{\zn/\gs\rightarrow\tau^+\tau^-}\xspace}

\newcommand{\tauptaum}{\ensuremath{\tau^+\tau^-}\xspace}
\newcommand{\qqbar  }{\ensuremath{\mathrm{q\bar{q}}}\xspace}
\newcommand{\ggtau  }{\ensuremath{\gamma^{\star}\gamma\to\tauptaum}\xspace}
\newcommand{\gs     }{\ensuremath{\gamma^\star}\xspace}
\newcommand{\gsg    }{\ensuremath{\gs\gamma}\xspace}
\newcommand{\gsgs   }{\ensuremath{\gs\gs}\xspace}

\newcommand{\gsgshad}{\ensuremath{\gs\gs\rightarrow\mbox{hadrons}}\xspace}
\newcommand{\chiq   }{\ensuremath{\chi^2}\xspace}
\newcommand{\dsdx   }{\ensuremath{\der\sigma/\der x}\xspace}
\newcommand{\dedx   }{\ensuremath{\der E/\der x}\xspace}
\newcommand{\fitres }{\ensuremath{\ftqn=(\Z{0.08}{0.02}{0.05}{0.03})%
           +(\Z{0.13}{0.01}{0.01}{0.01})\,\ln\,\qsq}\xspace}
\newcommand{\fitold }{\ensuremath{\ftqn=(\Z{0.16}{0.05}{0.17}{0.16})%
           +(\Z{0.10}{0.02}{0.05}{0.02})\,\ln\,\qsq}\xspace}
\newcommand{\po     }{\ensuremath{\phantom{0}}\xspace}
\newcommand{\pw     }{\ensuremath{\phantom{1}}\xspace}
\begin{document}
\begin{titlepage}
\begin{center}
{\Large  EUROPEAN ORGANIZATION FOR NUCLEAR RESEARCH}
\end{center}
\bigskip\bigskip
\begin{flushright}
 CERN-EP-2002-014 \\  15 February 2002 
\end{flushright}
\begin{center}
{\huge\bf\boldmath
 Measurement of the Hadronic \\
 Photon Structure Function \ft \\\vspace{0.25cm}
 at LEP2
 \unboldmath}
\end{center}
\begin{center}   
\end{center}
\begin{center}{\LARGE The OPAL Collaboration}\end{center}
%
%
\begin{abstract}
 The hadronic structure function of the photon \ftxq is measured 
 as a function of Bjorken $x$ and of the photon virtuality \qsq
 using deep-inelastic scattering data taken by the OPAL detector at LEP 
 at \epem centre-of-mass energies from 183 to 209~GeV. 
 Previous OPAL measurements of the $x$ dependence of \ft are extended
 to an average \qsq of $\qzm=780$~\gevsq using data in the kinematic range 
 $0.15<x<0.98$.
 The \qsq evolution of \ft is studied for $12.1 < \qzm < 780$~\gevsq 
 using three ranges of $x$.
 As predicted by QCD, the data show positive scaling violations in \ft 
 with \fitres, where \qsq is in \gevsq, for the central $x$ region 0.10--0.60.
 Several parameterisations of \ft are in qualitative agreement with the 
 measurements whereas the quark-parton model prediction fails to describe
 the data.
\end{abstract}
\bigskip\bigskip\bigskip
\begin{center} {\large (Submitted to Physics Letters B)}
\end{center}
\end{titlepage}
\begin{center}{\Large        The OPAL Collaboration
}\end{center}\bigskip
\begin{center}{
G.\thinspace Abbiendi$^{  2}$,
C.\thinspace Ainsley$^{  5}$,
P.F.\thinspace {\AA}kesson$^{  3}$,
G.\thinspace Alexander$^{ 22}$,
J.\thinspace Allison$^{ 16}$,
G.\thinspace Anagnostou$^{  1}$,
K.J.\thinspace Anderson$^{  9}$,
S.\thinspace Asai$^{ 23}$,
D.\thinspace Axen$^{ 27}$,
G.\thinspace Azuelos$^{ 18,  a}$
% ...
}\end{center}\bigskip
\bigskip
$^{  1}$School of Physics and Astronomy, University of Birmingham,
Birmingham B15 2TT, UK
\newline
$^{  2}$Dipartimento di Fisica dell' Universit\`a di Bologna and INFN,
I-40126 Bologna, Italy
\newline
$^{  3}$Physikalisches Institut, Universit\"at Bonn,
D-53115 Bonn, Germany
\newline
% ...
$^{  5}$Cavendish Laboratory, Cambridge CB3 0HE, UK
\newline
% ...
$^{  9}$Enrico Fermi Institute and Department of Physics,
University of Chicago, Chicago IL 60637, USA
\newline
% ...
$^{ 16}$Department of Physics, Schuster Laboratory, The University,
Manchester M13 9PL, UK
\newline
% ...
$^{ 18}$Laboratoire de Physique Nucl\'eaire, Universit\'e de Montr\'eal,
Montr\'eal, Quebec H3C 3J7, Canada
\newline
% ...
$^{ 22}$Department of Physics and Astronomy, Tel Aviv University,
Tel Aviv 69978, Israel
\newline
$^{ 23}$International Centre for Elementary Particle Physics and
Department of Physics, University of Tokyo, Tokyo 113-0033, and
Kobe University, Kobe 657-8501, Japan
\newline
% ...
$^{ 27}$University of British Columbia, Department of Physics,
Vancouver BC V6T 1Z1, Canada
\newline
% ...
$^{  a}$ and at TRIUMF, Vancouver, Canada V6T 2A3
\newline
% ...
%
%
\section{Introduction}
\label{sec:intro}
 Much of the present knowledge of the structure of the photon has been
 obtained from measurements of the photon structure function \ft in
 deep-inelastic electron-photon\footnote{For conciseness positrons are 
 also referred to as electrons.} scattering at \epem colliders,
 see~\cite{NIS-9904} for a recent review.
 The large statistics and high electron energies of the full LEP2 
 programme permit the extension of the measurement of \ft to higher values 
 of \qzm than have been probed at LEP1.
 The photon structure function \ft is expected to increase only 
 logarithmically with \qsq~\cite{GRI-7201LIP-7501ALT-7701DOK-7701}.
 Therefore, the large range of \qsq values accessible at LEP, which
 extends from about 1~\gevsq to several thousand~\gevsq, makes it an ideal 
 place to study the evolution.
 \par
 The measurement of \ft in \epem interactions
 is based on the deep-inelastic electron-photon scattering reaction,
 ${\rm e}(k)\,\gamma(p)\rightarrow{\rm e}(k')\,+\,\mbox{hadrons}$, 
 proceeding via the exchange of a virtual photon, $\gamma^*(q)$, where the
 symbols in brackets denote the four-momentum vectors of the particles.
 The flux of quasi-real photons can be calculated using the
 equivalent photon approximation~\cite{WEI-3401WIL-3401BUD-7501}.
 The cross-section for deep inelastic electron-photon scattering
 is expressed as:
%
 \begin{equation}
  \frac{{\rm d}^2\sigma_{\rm e\gamma\rightarrow {\rm e X}}}{{\rm d}x{\rm d}Q^2}
 =\frac{2\pi\aemsq}{x\,Q^{4}}
  \left[\left( 1+(1-y)^2\right) \ftxq - y^{2} \flxq\right]
 \label{eqn:Xsect}
 \end{equation}
%
 where $\qsq=-q^2$.
 The usual dimensionless variables of deep inelastic
 scattering, $x$ and $y$, are defined as $x=\qsq/2(p\cdot q)$
 and $y=(p\cdot q)/(p\cdot k)$, and \aem is the fine structure constant.
 The structure function \ft is related to the charge-weighted sum of
 the parton densities of the photon (see {\it e.g.}~\cite{NIS-9904}).
 In the kinematic region of low values of $y$ studied ($y^2\ll 1$) 
 the contribution of the term proportional to the longitudinal structure 
 function \flxq is negligible~\cite{NIS-9904}.
 \par
 The analysis presented here is based on 632~\invpb of data at \epem 
 centre-of-mass energies \ssee of 183 to 209~GeV, with a luminosity 
 weighted average of $\ssee=197.1$~GeV, taken by the OPAL 
 experiment in the years 1997--2000.
 It extends the measurements of \ft as a function of $x$ 
 up to $\qzm=780$~\gevsq, and significantly improves on the 
 precision of the measurement of the \qsq evolution of \ft.
 This analysis not only tests perturbative QCD but also measures \ft
 at large \qsq, a previously unexplored region in \epem collisions.
 This is approximately the region which has also been probed in jet production
 at HERA~\cite{ZEU-9901,NIS-0101}.
 \par
 The paper is organised as follows. 
 After the description of the OPAL detector in Section~\ref{sec:detec} the
 data selection is detailed in Section~\ref{sec:evsel}, followed by the 
 description of the Monte Carlo simulation and background estimates in 
 Section~\ref{sec:MC}.
 The results are presented in Section~\ref{sec:resu}.
 These comprise: the quality of the description of the observed hadronic
 final state by the Monte Carlo models, Section~\ref{sec:comp};
 the measurement of \ft at high \qsq, Section~\ref{sec:highq2}; and
 the measurement of the \qsq evolution of \ft, Section~\ref{sec:evol}.
 Conclusions are given in Section~\ref{sec:concl}.
%
%
\section{The OPAL detector}
\label{sec:detec}
 A detailed description of the OPAL detector can be found 
 in~\cite{OPALPR021}, and therefore only a brief account of the main 
 features relevant to the present analysis will be given here.
 \par
 The central tracking system is located inside a solenoidal magnet which
 provides a uniform axial magnetic field of 0.435~T along the beam
 axis\footnote{In the OPAL coordinate system the $x$ axis points
 towards the centre of the LEP ring, the $y$ axis upwards and
 the $z$ axis in the direction of the electron beam. In this paper the
 polar angle $\theta$ is defined with respect to the closest orientation 
 of the $z$ axis.}.
 The magnet is surrounded by a lead-glass electromagnetic
 calorimeter (ECAL) and a hadronic sampling calorimeter (HCAL).
 Outside the HCAL, the detector is surrounded by muon
 chambers. There are similar layers of detectors in the
 endcaps. The region around the beam pipe on both sides
 of the detector is covered by the forward calorimeters and the
 silicon-tungsten luminometers.
 \par
 Starting with the innermost components, the tracking system consists of 
 a high precision silicon microvertex detector~\cite{ALL-9401ALL-9301}, 
 a precision vertex drift chamber, a large-volume jet chamber with 159 layers
 of axial anode wires, and a set of $z$ chambers used to improve the 
 measurement of the track coordinates along the beam direction.
 The transverse momenta \pt of tracks with respect to the $z$ direction of 
 the detector are measured with a precision of
 $\sigma_{\pt}/\pt=\sqrt{0.02^2+(0.0015\cdot \pt)^2}$ (\pt in GeV)
 in the central region,
 $\theta>753$~mrad.
 The jet chamber also provides energy loss, \dedx, 
 measurements which are used for particle identification.
 \par
 The ECAL covers the complete azimuthal range for polar angles
 that satisfy 
 $\theta>200$~mrad.
 The barrel section, which covers the range 
 $\theta>609$~mrad,
 consists of a cylindrical array of 9440 lead-glass blocks with a depth of
 $24.6$ radiation lengths. The endcap sections (EE) consist of 1132 
 lead-glass blocks with a depth of more than $22$ radiation lengths, 
 covering angles in the range
 $200< \theta < 609$~mrad.
 The electromagnetic energy resolution of the EE calorimeter is about
 $15\%/\sqrt{E}$~($E$ in GeV) at polar angles above 350~mrad, but 
 deteriorates closer to the edge of the detector.
 \par
 The forward calorimeters (FD) at each end of the OPAL detector
 consist of cylindrical lead-scintillator calorimeters with a depth of
 24 radiation lengths divided azimuthally into 16 segments.
 The electromagnetic energy resolution of the FD calorimeter is about
 $18\%/\sqrt{E}$~($E$ in GeV).
 The clear acceptance of the forward calorimeters covers the range
 $60< \theta < 140$~mrad.
 Three planes of proportional tube chambers at 4 radiation lengths
 depth in the calorimeter measure the directions of 
 showers with a precision of approximately 1~mrad.
 \par
 The silicon tungsten detectors (SW)~\cite{AND-9401} at each end of the 
 OPAL detector lie in front of the forward calorimeters.
 Their clear acceptance covers a polar angular region between 33 and 59~mrad.
 Each SW calorimeter consists of 19 layers of silicon detectors and 18
 layers of tungsten, corresponding to a total of 22 radiation
 lengths. Each silicon layer consists of 16 wedge-shaped 
 silicon detectors. The electromagnetic energy resolution is about
 $25\%/\sqrt{E}$ ($E$ in GeV). The radial position of electron
 showers in the SW calorimeter can be determined with a
 typical resolution of 0.06 mrad in the polar angle $\theta$.
%
%
\section{Kinematics and data selection}
\label{sec:evsel}
 The interactions of two photons are classified according to the 
 virtualities of the photons. For this analysis photons with a virtuality
 of less than 4.5~\gevsq are called quasi-real photons, $\gamma$,
 and the other photons are virtual photons, \gs.
 As a shorthand, events caused by the interactions of the three possible 
 combinations are called \GG, \gsg and \gsgs events.
 \par
 To measure \ftxq, the distribution of \gsg events in $x$ and \qsq is needed. 
 These variables are related to the experimentally measurable quantities
 $W$, \etag and \ttag by
%
\begin{equation}
  \qsq = 2\,\eb\,\etag\,(1- \cos\ttag)
  \label{eqn:qsq}
\end{equation}
%
 and
%
\begin{equation}
  x = \frac{\qsq}{\qsq+W^2+\psq}
  \label{eqn:Xcalc}
\end{equation}
%
 where \eb is the energy of the beam electrons, \etag and \ttag are the 
 energy and polar angle of the deeply inelastically scattered (or `tagged') 
 electron, \wsq is the invariant mass squared of the hadronic final state,
 and $\psq=-p^2$ is the negative value of the virtuality of the
 quasi-real photon.
 The requirement that the electron associated with the quasi-real photon 
 is not seen in the detector (anti-tag condition) ensures that 
 $\psq \ll \qsq$, so \psq is neglected when calculating $x$ from 
 Equation~\ref{eqn:Xcalc}.
 The electron mass is neglected throughout. 
 \par
 Three samples of \gsg events are studied in this analysis, classified
 according to the subdetector in which the scattered electron is observed. 
 Electrons are measured using the SW, FD and EE detectors.
 Events are selected by applying cuts on the scattered electrons and on 
 the hadronic final state.
 A scattered electron is selected by requiring $\etag \ge 0.75/0.75/0.70\,\eb$
 and polar angles $33.25/60/230\le\ttag\le 55/120/500$~mrad
 for the SW/FD/EE samples.
 For the SW sample the energy cut effectively eliminates events originating
 from random coincidences between
 off-momentum\footnote{Off-momentum electrons originate from beam gas
 interactions  far from the OPAL interaction region and are 
 deflected into the detector by the focusing quadrupoles.} 
 beam electrons faking a scattered electron and untagged \GG 
 events~\cite{OPALPR344}.
 For the EE sample special measures have to be taken to avoid fake electron
 candidates.
 To remove electron candidates originating from energetic electromagnetic
 calorimeter clusters stemming {\it e.g.}~from hadronic final states in the
 reaction \znhad, an isolation cut is applied which requires that less 
 than 3~\gev is deposited in a cone of 500~mrad half-angle around the electron 
 candidate (electron isolation cut).
 \par
 To ensure that the virtuality of the quasi-real photon is small, the highest
 energy electromagnetic cluster in the hemisphere opposite to the one 
 containing the scattered electron must have an energy $\ea \le 0.25\,\eb$
 (the anti-tag condition).
 To reject background from deep-inelastic scattering events with leptonic
 final states, the number of tracks in the event passing quality
 cuts~\cite{OPALPR150} and originating from the hadronic final state, 
 \nch, must be at least three/three/four for the SW/FD/EE samples,
 of which at least two tracks must not be identified as electrons,
 based on the energy-loss measurement in the jet chamber. 
 The tracks and the calorimeter clusters are reconstructed using 
 standard OPAL techniques~\cite{OPALPR150} which avoid double counting
 the energy of particles that produce both tracks and clusters.
 The visible invariant mass \Wvis of the hadronic system is calculated
 from tracks and calorimeter clusters, including contributions from 
 energy measured in the SW and FD calorimeters.
 For the EE sample, because of the high probability that the scattered 
 electron will shower in the dead material (ranging from $2-6$ radiation 
 lengths) in front of the EE calorimeter,
 energy deposits close to the electron are likely to belong to the electron.
 Therefore, for this sample, all tracks and clusters within a cone 
 of 200~mrad half-angle about the direction of the electron candidate 
 are excluded from the calculation of \Wvis.
 To remove the region dominated by resonance production and to reject the 
 background from \znhad, the measured \Wvis is required to be in the range 
 $2.5 < \Wvis < 60/60/50$~\gev for the SW/FD/EE samples.
 The stronger cut on \Wvis applied to the EE sample reflects the fact that
 the background from \znhad is larger for this sample than for the other 
 samples.
 \par 
 The cuts applied to each sample are listed in Table~\ref{tab:cuts}.
 The numbers of events in each sample passing the cuts are listed in 
 Table~\ref{tab:samples}, together with the numbers of signal events after 
 subtracting the background contributions described below.
 Trigger efficiencies were evaluated from the data using sets of 
 separate triggers, and were found to be larger than $99\%$ for 
 events within the selection cuts.
%
%
\section{Monte Carlo simulation and background estimation}
\label{sec:MC}
 Monte Carlo programs are used to simulate signal events and to provide 
 background estimates. 
 All Monte Carlo events are passed through the OPAL detector 
 simulation~\cite{ALL-9201} and the same reconstruction and analysis 
 chain as used for real events.
 \par
 The Monte Carlo generators used to simulate signal events are
 HERWIG 5.9+\kt(dyn)~\cite{MAR-9201}, PHOJET 1.05~\cite{ENG-9501ENG-9601}
 and the Vermaseren program~\cite{VER-7901VER-8301}.
 The main reason for using a second Monte Carlo together with HERWIG
 is to have an additional model that contains different assumptions for 
 modelling the hard scattering and the hadronisation process.
 HERWIG is a general purpose Monte Carlo program which includes
 deep inelastic electron-photon scattering. 
 The HERWIG 5.9+\kt(dyn) version uses a modified transverse momentum, \kt,
 distribution for the quarks inside the photon for hadron-like events.
 The upper limit of the \kt distribution is dynamically (dyn) adjusted 
 according to the hardest scale in the event, which is of order \qsq.
 This version was found to better describe the observed hadronic final 
 states in three of the LEP experiments~\cite{OPALPR316} than the 
 original version HERWIG 5.9.
 In HERWIG the cluster model is used for the hadronisation process.
 PHOJET simulates hard interactions through perturbative QCD and 
 soft interactions through Regge phenomenology, and the hadronisation is
 modelled by JETSET~\cite{SJO-0101}.
 Since it is recommended by the authors to use PHOJET only for \qsq values
 smaller than about 50~\gevsq, the Vermaseren model is used for the EE sample.
 The Vermaseren program is based on the quark-parton model (QPM) and
 the quark masses assumed in the event generation are 0.325~GeV for $u,d,s$
 and 1.5~GeV for c quarks. 
 For each Monte Carlo sample the generated integrated luminosity
 is at least 10 times that of the data.
 \par
 The HERWIG and PHOJET samples were generated using the leading order
 GRV~\cite{GLU-9201GLU-9202} parameterisation of \ft, taken from the 
 PDFLIB library~\cite{PLO-9301}, as the input structure function. 
 This version assumes massless charm quarks.
 Since PHOJET is not based on the cross-section formula for deep 
 inelastic electron-photon scattering, the program always produces the 
 same $x$ and \qsq distributions independent of the input structure function.
 Therefore the $x$ distribution of PHOJET was reweighted to match that from 
 HERWIG, as described in~\cite{OPALPR314}.
 This is not a strong limitation, because the main emphasis lies on the 
 alternative hadronisation model. The result of the unfolding procedure 
 is expected to be almost independent of the actual underlying $x$ 
 distribution of the Monte Carlo sample used. 
 The numbers of expected signal events from the HERWIG program are 
 listed in Table~\ref{tab:samples}.
 \par
 For the SW and FD samples the dominant background comes from the reaction
 $\epem\rightarrow\epem\tauptaum$ proceeding via the multiperipheral 
 diagram~\cite{NIS-9904}.
 This was simulated using the Vermaseren program.
 In contrast, for the EE sample the dominant background stems from the 
 reaction \znhad, which was simulated using PYTHIA~\cite{SJO-9401SJO-9301}.
 The next largest backgrounds are $\epem\rightarrow\epem\tauptaum$ followed
 by non-multiperipheral four-fermion events
 with $\ee\qqbar$ final states (denoted by 4-fermion eeqq), which were 
 simulated with GRC4f~\cite{FUJ-9801}, and \zntau, which was simulated with 
 the KK~\cite{JDH-0001} program.
 Because the aim is to measure the structure function of the quasi-real 
 photon, events stemming from the interaction of two virtual photons with
 hadronic final states are also treated as background.
 For the SW and FD samples these were generated using PHOJET 1.10 with the 
 virtualities of both photons restricted to be above 4.5~\gevsq.
 For the EE sample they have been estimated using the Vermaseren program.
 The contribution to the background due to all other Standard Model processes
 was found to be negligible in all the samples.
 The numbers of events from the dominant 
 background sources for each data sample are listed in Table~\ref{tab:samples}.
%
%
\section{Results}
\label{sec:resu}
\subsection{Comparison of data and Monte Carlo}
\label{sec:comp}
\begin{figure}[tbp]
\begin{center}
{\includegraphics[width=1.0\linewidth]{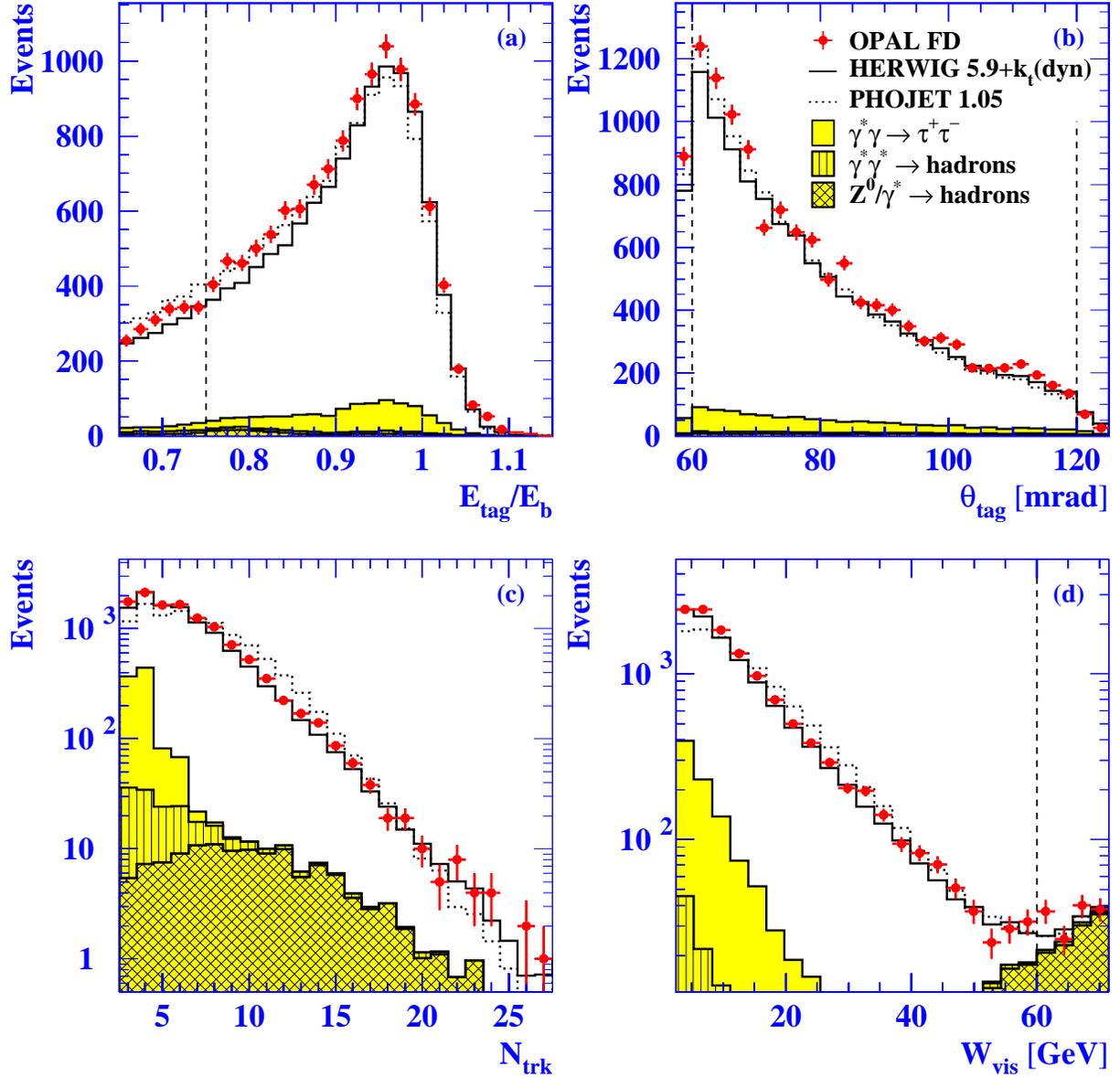}}
\caption{Comparison of data distributions for the FD sample 
         with Monte Carlo predictions.
         The open histograms are the sum of the signal prediction and
         the contributions of the major background sources, shown both 
         for the HERWIG (full lines) and PHOJET (dotted lines) models.
         The Monte Carlo predictions are normalised
         to the data luminosity, except for 
         PHOJET where the sample has been normalised such that the predicted 
         number of events for the SW sample is the same as that of HERWIG.
         All selection cuts have been applied, except
         for any cut on the variable in the plot
         (indicated as dashed lines if within the region shown). 
         The distributions are:
         (a) \etag/\eb, the energy of the scattered electron as a fraction
             of the energy of the beam electrons;
         (b) \ttag, the polar angle of the scattered electron;
         (c) \nch, the number of tracks originating from the hadronic
             final state; and
         (d) \Wvis, the measured invariant mass of the hadronic final state.
         The errors given are only statistical.
        }\label{fig:fig01}
\end{center}
\end{figure}

\begin{figure}[tbp]
\begin{center}
{\includegraphics[width=1.0\linewidth]{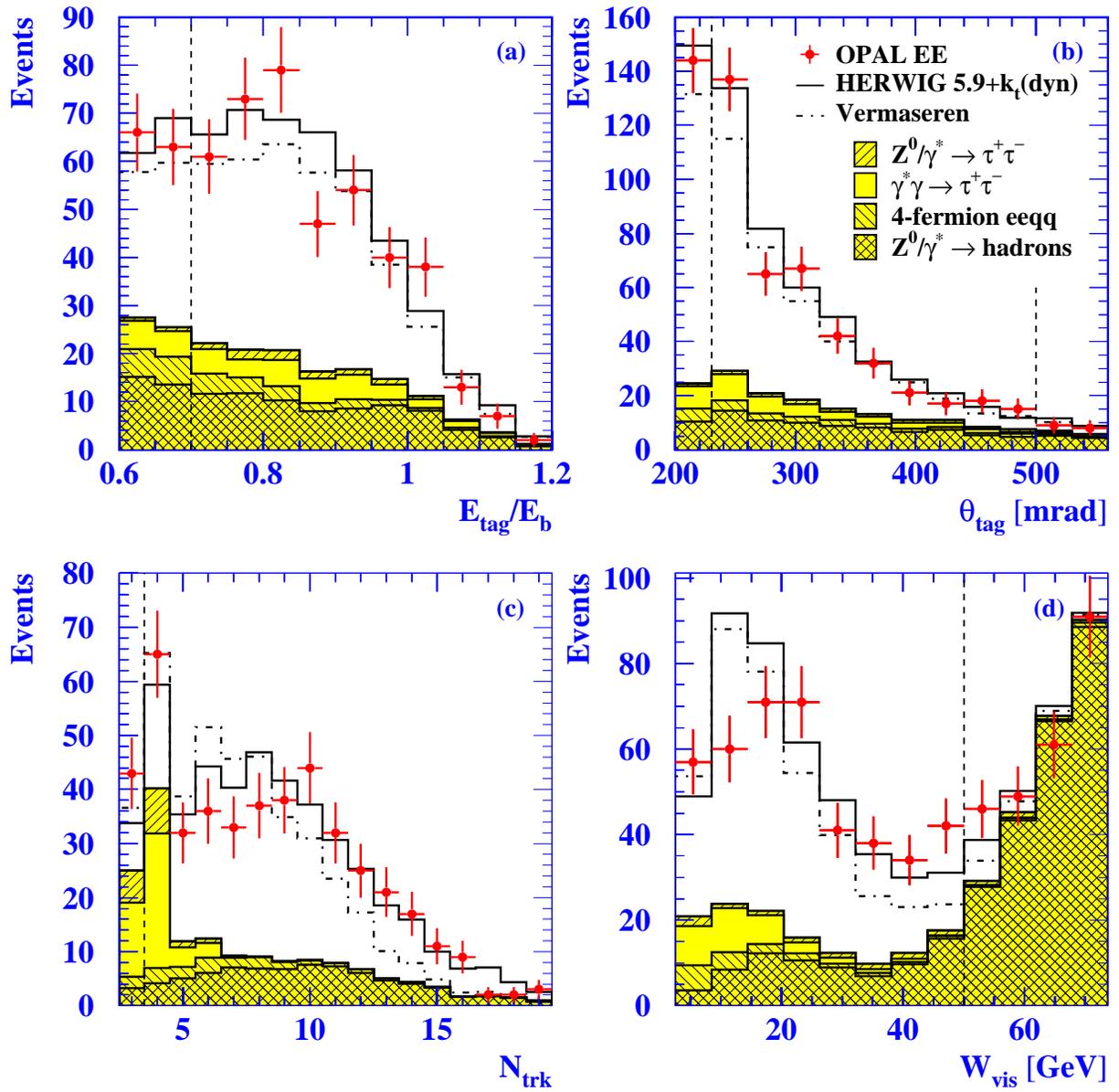}}
\caption{Comparison of data distributions for the EE sample with the
         Monte Carlo predictions. 
         See Figure~\protect\ref{fig:fig01} for details. 
        }\label{fig:fig02}
\end{center}
\end{figure}

 Monte Carlo samples are used in an unfolding procedure to extract
 the differential cross-section \dsdx and \ftxq from the data. 
 Therefore, apart from explicit effects due to the variation of the structure 
 function, a good description of the data distributions by the Monte Carlo 
 is needed both for electron variables, which are used to measure \qsq, and
 for hadronic variables, which determine \wsq.
 The analysis of the SW sample closely follows that presented
 in~\cite{OPALPR314} but includes three times the data integrated luminosity.
 The quality of the description of this sample is similar to that
 presented in~\cite{OPALPR314}.
 The analysis of the FD and EE samples at LEP2 energies is new.
 Figures~\ref{fig:fig01} and~\ref{fig:fig02} show comparisons between data
 and Monte Carlo distributions for these two samples.
 The quantities shown are
 (a) \etag/\eb, the energy of the scattered electron as a fraction
     of the energy of the beam electrons, 
 (b) \ttag, the polar angle of the scattered electron,
 (c) \nch, the number of tracks originating from the hadronic
     final state, and
 (d) \Wvis, the measured invariant mass of the hadronic final state.
 The FD sample, Figure~\ref{fig:fig01}, is compared to the Monte Carlo 
 prediction of the HERWIG and PHOJET (without reweighting of the $x$ 
 distribution) signal events together with background estimates.
 The PHOJET sample has been normalised such that the predicted 
 number of events for the SW sample is the same as that of HERWIG, 
 so the PHOJET distributions only allow for a shape comparison.
 The HERWIG Monte Carlo model predicts slightly fewer events than are 
 observed in the data and, in general, the shapes of the data distributions 
 are better described by HERWIG than by PHOJET.
 The EE sample, Figure~\ref{fig:fig02}, is compared to the Monte Carlo 
 prediction of the HERWIG and Vermaseren signal events together with
 background estimates.
 The data distributions of the energy and polar angle of the scattered 
 electron are well described by the Monte Carlo predictions.
 For the variables related to the hadronic final state there are 
 apparent differences in shape.
 \par
 The hadronic energy flow for the SW sample has been studied
 in~\cite{OPALPR314}.
 On average, about 5$\%$ of the energy is deposited in SW, and about 20-25$\%$ 
 in FD and SW combined. The numbers for the FD and EE sample are even lower.
 It was verified  that scaling the energy in the forward region has a small 
 impact on the measured \ft for $x>0.1$.  Consequently, in the present 
 analysis this energy is not scaled.
 \par
\begin{figure}[tbp]
\begin{center}
{\includegraphics[width=1.0\linewidth]{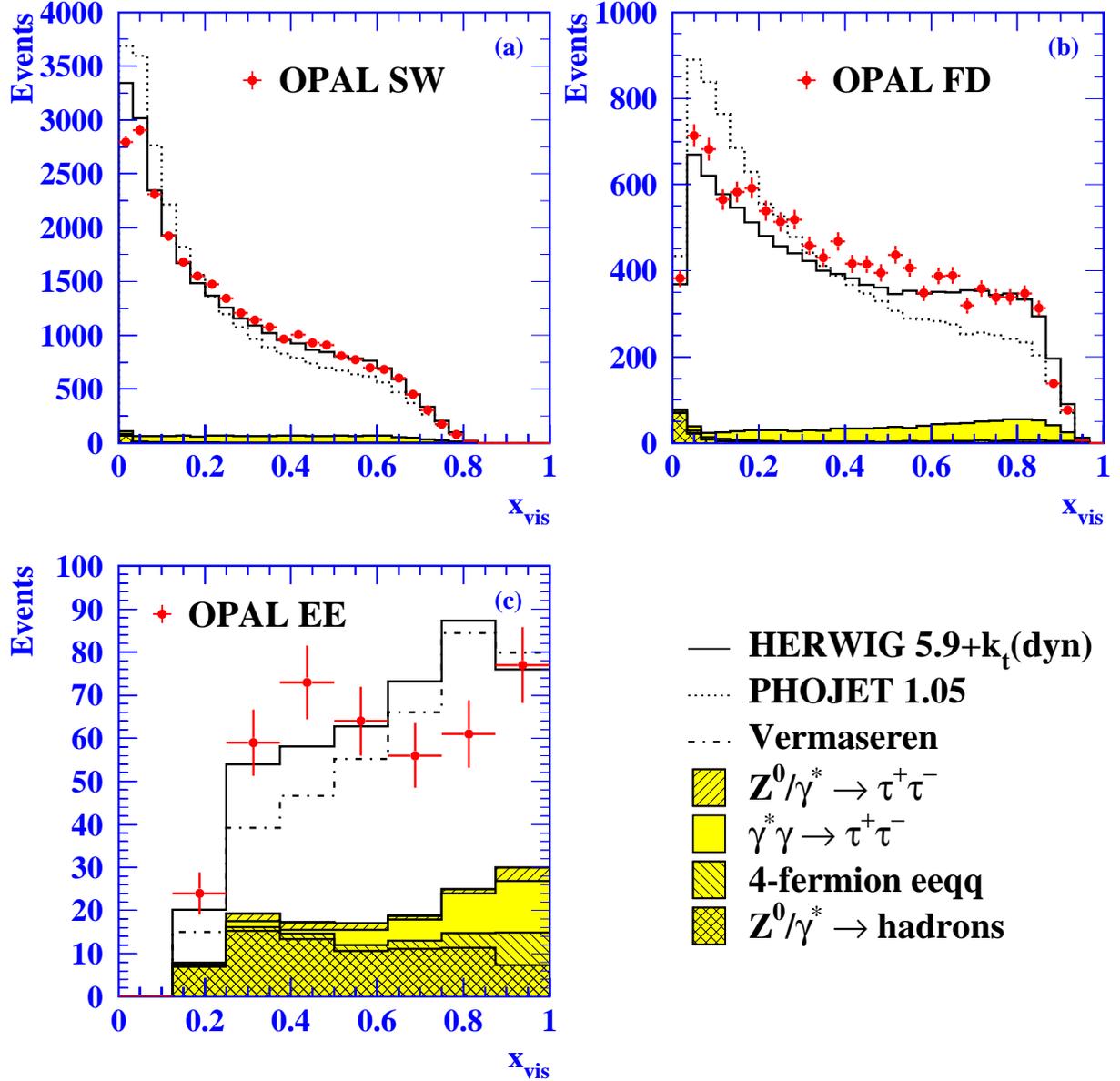}}
\caption{The measured \xvis distributions for the three samples
         (a) SW, (b) FD and (c) EE. 
         The data are compared to Monte Carlo predictions
         containing signal and background contributions normalised
         to the data luminosity, except for 
         PHOJET where the sample has been normalised such that the predicted 
         number of events for the SW sample is the same as that of HERWIG.
         The errors given are only statistical.
        }\label{fig:fig03}
\end{center}
\end{figure}

 The quantity \xvis, obtained from \qsq and \Wvis, is shown in
 Figure~\ref{fig:fig03} for the three samples.
 It should be noted that for the SW sample, for $\xvis>0.1$, the HERWIG model 
 using the GRV parameterisation of \ft qualitatively follows the data, which 
 means that the \ft found from the data should be similar to the expectation
 from GRV.
 In contrast, for the FD sample for $0.1<\xvis<0.7$ the HERWIG prediction
 is systematically lower than the data.
 Due to the shortcoming of the PHOJET model discussed above, the 
 description of the \xvis distribution is unsatisfactory when using the
 PHOJET model without reweighting of the $x$ distribution.
 For the EE sample the difference in shape of the \Wvis distribution is 
 reflected in the observed difference between the data and both Monte Carlo
 models for the \xvis distribution.
%
%
\subsection{Measurement of \boldmath\dsdx and \ft at high \qsq \unboldmath}
\label{sec:highq2}
 The differential cross-section \dsdx and the structure function
 \ft are obtained from the data by unfolding the \xvis distribution of 
 the EE sample, after applying additional cuts on \qsq.
 The main problem in measurements of \ft at low $x$, i.e.~$x<0.1$,
 is the dependence of \ft 
 on the Monte Carlo modelling, which enters when the unfolding process is used
 to relate the visible distributions to the underlying $x$ distribution.
 This problem is less severe at medium to large values of $x$, in particular
 for the high \qsq EE sample, where the hadronic final state has much more 
 transverse momentum and as a consequence is better contained in the detector. 
 Therefore the correlation between the measured invariant mass \Wvis and
 the true $W$, {\it e.g.}~as given by HERWIG, is much better at large $x$,
 so the results can be expected to have a smaller
 dependence on the Monte Carlo modelling of the hadronic final state.
 \par
 No attempt has been made in this analysis to access the region of $x<0.1$, so
 using a one dimensional unfolding on a linear scale in $x$ is appropriate,
 in contrast with~\cite{OPALPR314}. 
 For this purpose the RUN program~\cite{BLO-8401BLO-9601} has been used. 
 Technically, RUN uses a set of Monte Carlo events which are based on an
 input \ftxq and carry the information about the correlation of \xvis and $x$.
 A continuous weight function is defined which depends only on $x$. 
 This function is constructed from individual weight factors for each 
 Monte Carlo event.
 These weight factors are obtained by fitting the \xvis distribution
 of the Monte Carlo sample to the measured \xvis distribution of the data,
 such that the reweighted Monte Carlo events describe the \xvis distribution
 of the data as well as possible.
 After the unfolding the two \xvis distributions are consistent.
 The unfolded \ftxq from the data is then obtained by multiplying the input
 \ftxq of the Monte Carlo with the weight function.
 For further details the reader is referred to~\cite{NIS-9904}.
 It has been demonstrated in~\cite{OPALPR314} that this procedure is 
 independent of the input structure function used in the Monte Carlo.
 \par
 Radiative corrections and the dependence of \ftxqp on \psq 
 are treated as in the previous OPAL analysis~\cite{OPALPR314}.
 The radiative corrections applied to the data have been estimated using 
 the RADEG program~\cite{LAE-9603LAE-9701}.
 They are obtained for each bin in $x$ and \qsq using the 
 SaS1D~\cite{SCH-9501} prediction of \ft.
 No correction for the effect of non-zero \psq has been made,
 see Refs.~\cite{OPALPR314,NIS-9904} for further details.
 The average value of \psq of the data samples as predicted by the HERWIG
 program is about 0.2~\gevsq. Note however that HERWIG does not take into 
 account the \psq dependence of \ft.
 \par
\begin{figure}[tbp]
\begin{center}
{\includegraphics[width=1.0\linewidth]{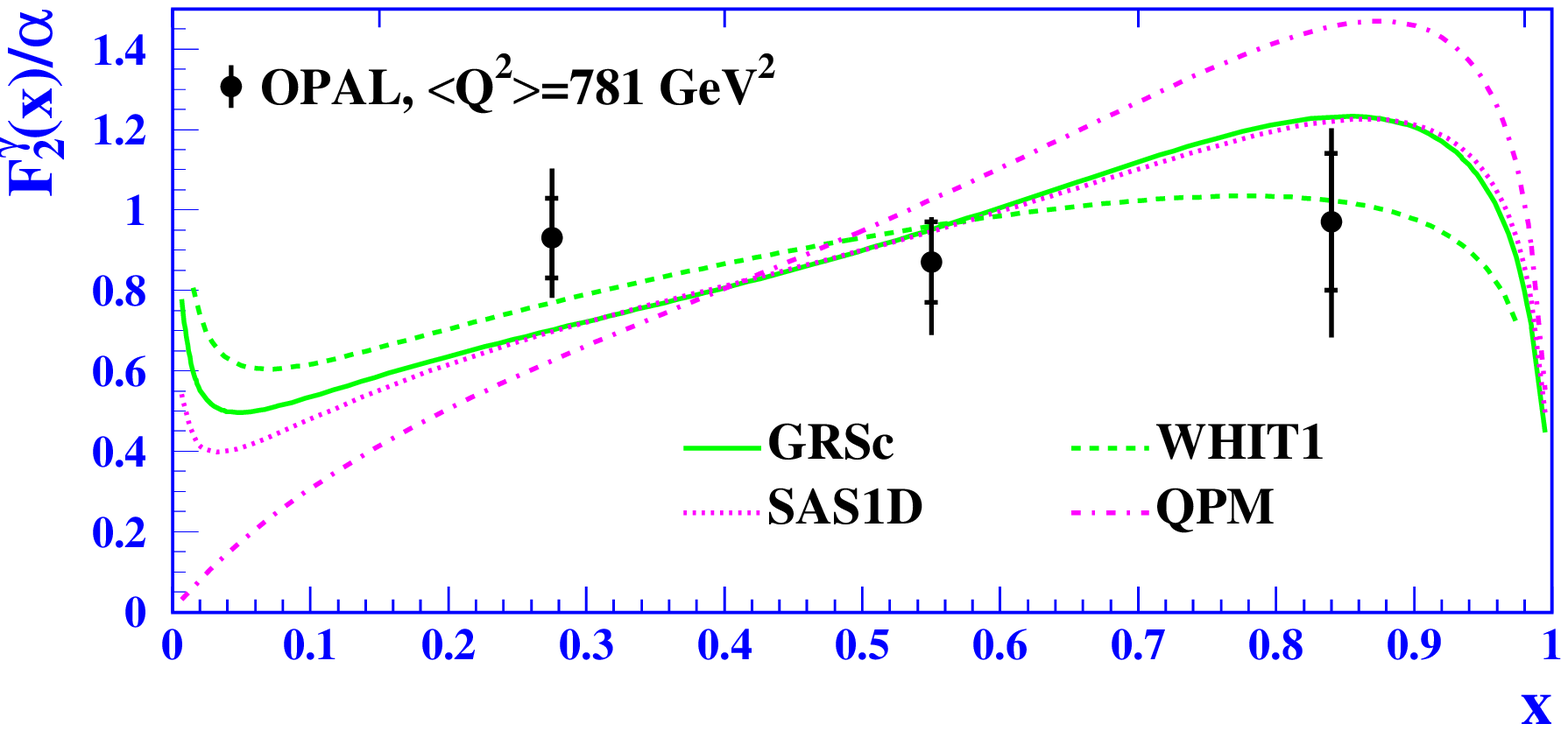}}
\caption{The measured \ftn as a function of $x$ for the EE sample. 
         The data are unfolded for $\qzm=780$~\gevsq and compared
         to the leading order predictions from the GRSc (full line), 
         SaS1D (dotted line), WHIT1 (dashed line) and QPM (dot-dashed line) 
         parameterisations of \ftn.
         The inner error bars represent the statistical errors and the outer 
         error bars represent statistical and systematic errors added in
         quadrature.
         The tick marks at the top of the figure represent the bin boundaries.
        }\label{fig:fig04}
\end{center}
\end{figure}

 After subtraction of background, the EE sample has been unfolded using three
 bins in $x$ spanning the range $0.15-0.98$ and for $400 < \qsq < 2350$~\gevsq.
 The central values are obtained using HERWIG as the input Monte Carlo
 model for the unfolding.
 Each data point is corrected for radiative effects as described above. 
 Bin-centre corrections are also applied as given by the average of the 
 GRSc~\cite{GLU-9902}, SaS1D and WHIT1~\cite{HAG-9501}
 predictions for the correction from the average \ft over the bin to the
 value of \ft at the nominal $x$ position.
 The result for \ftn is shown in Figure~\ref{fig:fig04} and listed in 
 Table~\ref{tab:results} together with the correlation matrix.
 In each bin of $x$ the result for \dsdx is also listed.
 The \dsdx values are corrected to the phase space given by the \qsq 
 range and $y<0.3$.
 \par
 Systematic errors are estimated by repeating the unfolding with one parameter
 varied at a time and determining the shift in the result. The systematic 
 errors are combined by adding all individual contributions in quadrature 
 separately for positive and negative contributions.
 The systematic effects considered for the EE sample are:
%
\begin{Enumerate}
\item Model dependence:\\
      The dependence on the Monte Carlo model used in the unfolding
      has been estimated by repeating the unfolding using the Vermaseren 
      sample and taking the full difference as the systematic error, both 
      for the positive and negative error.
\item Variations of cuts:\\
      The composition of the selected events was varied by
      changing the cuts one at a time.
      The size of the variations reflect the resolution of the measured 
      variables and the description of the data by the Monte Carlo models
      around the cut values.
      The variations are sufficiently small not to change the average \qsq
      of the sample significantly.
      The variations made are listed in Table~\ref{tab:cuts}.
\item Unfolding parameters:\\
      The number of bins used for the measured variable can be different from
      the number used for the true variable. The standard result has 5 bins in
      the measured variable. This was in turn reduced to 4 and increased to 6 
      to estimate the systematic effects of the unfolding.
\item Calibration of the tagging detector: \\
      The energy of the scattered electron in the Monte Carlo samples was 
      conservatively scaled by $\pm 1\%$~\cite{OPALPR286}.
\item Measurement of the hadronic energy:\\
      The main uncertainty is in the calibration of the response of the 
      electromagnetic calorimeter to hadronic energy for low energy 
      particles in the hadronic final state. The absolute energy 
      scale was varied by $\pm3\%$~\cite{OPALPR278} in the Monte Carlo 
      samples. 
\item Background modelling:\\
      To quantify the uncertainty on the most dominant background, stemming
      from the reaction \znhad, the KK program along with cluster 
      fragmentation from HERWIG has been used instead of PYTHIA with 
      string fragmentation.
\item Cone size for the \Wvis calculation:\\
      The size of the exclusion cone for the \Wvis calculation of 200~mrad 
      half-angle about the direction of the scattered electron has been varied
      by $\pm 30$~mrad.
\end{Enumerate}
%
 The size of the contributions to the error from the individual sources is 
 similar and no single source is dominant.
 When combining all error sources, the total estimated systematic error is 
 of the same order as the statistical error.
 \par
 The measured \ftn, shown in Figure~\ref{fig:fig04} together with
 several theoretical calculations, exhibits a flat behaviour.
 The leading order parameterisations of \ft from GRSc,
 SaS1D and WHIT1, which all include a contribution from massive charm quarks,
 are described in detail in~\cite{NIS-9904}. 
 The contribution from bottom quarks is negligible. 
 It can be seen that in this high \qsq regime the differences between these
 predictions are moderate, particularly in the central $x$-region.
 All these predictions are compatible with the data to within about 20$\%$, 
 with the WHIT1 parameterisation, which predicts the flattest behaviour,
 being closest to the data.
 The QPM curve, which models only the
 point-like component of \ft, is calculated for four active flavours with 
 masses of 0.325~\gev for light quarks and 1.5~\gev for charm quarks. 
 This prediction shows a much steeper behaviour in $x$ and is 
 disfavoured by the data.
%
%
\subsection{Measurement of the \boldmath \qsq evolution of \ft \unboldmath}
\label{sec:evol}
\begin{figure}[tbp]
\begin{center}
{\includegraphics[width=0.79\linewidth]{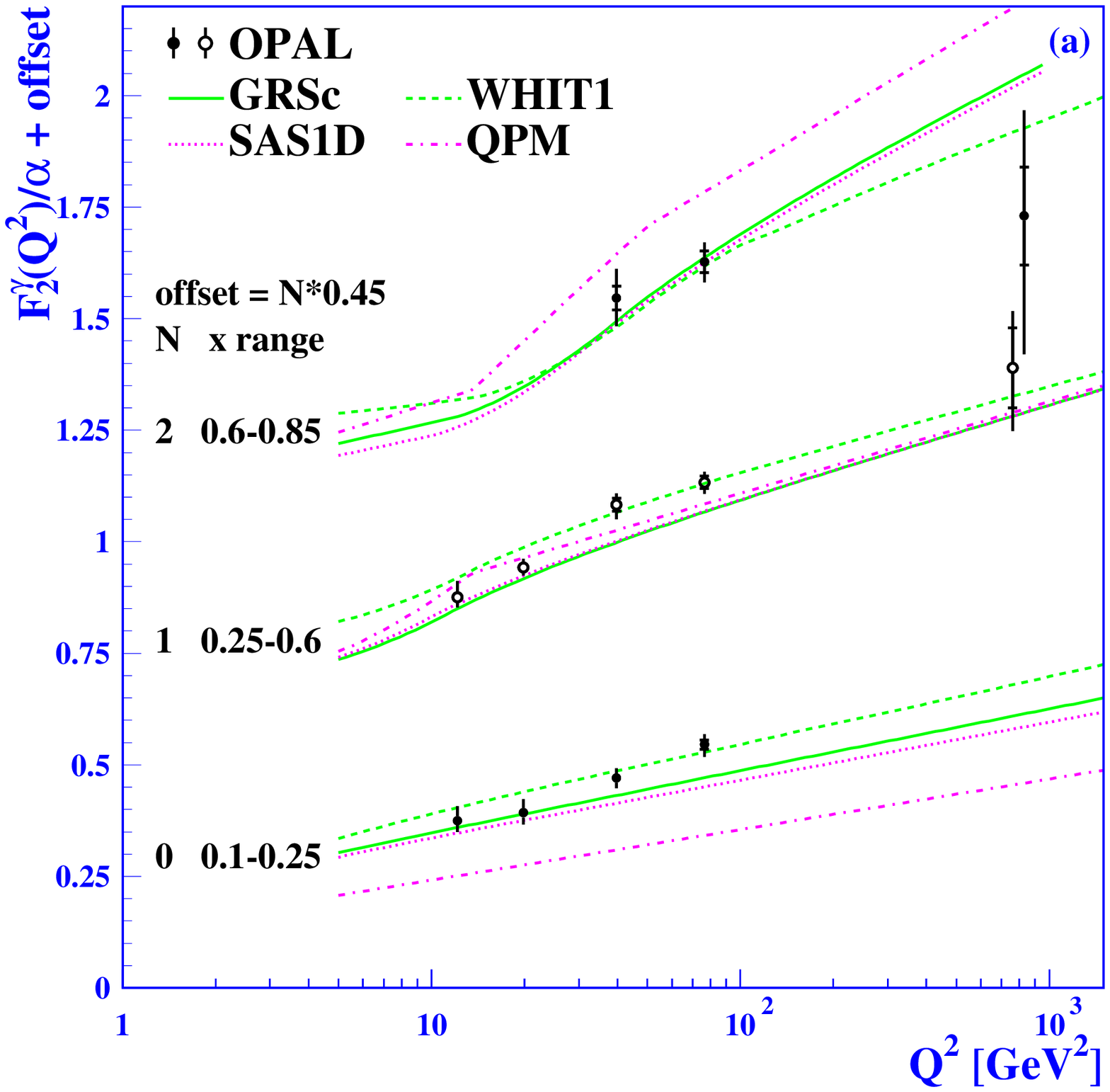}}
{\includegraphics[width=0.79\linewidth]{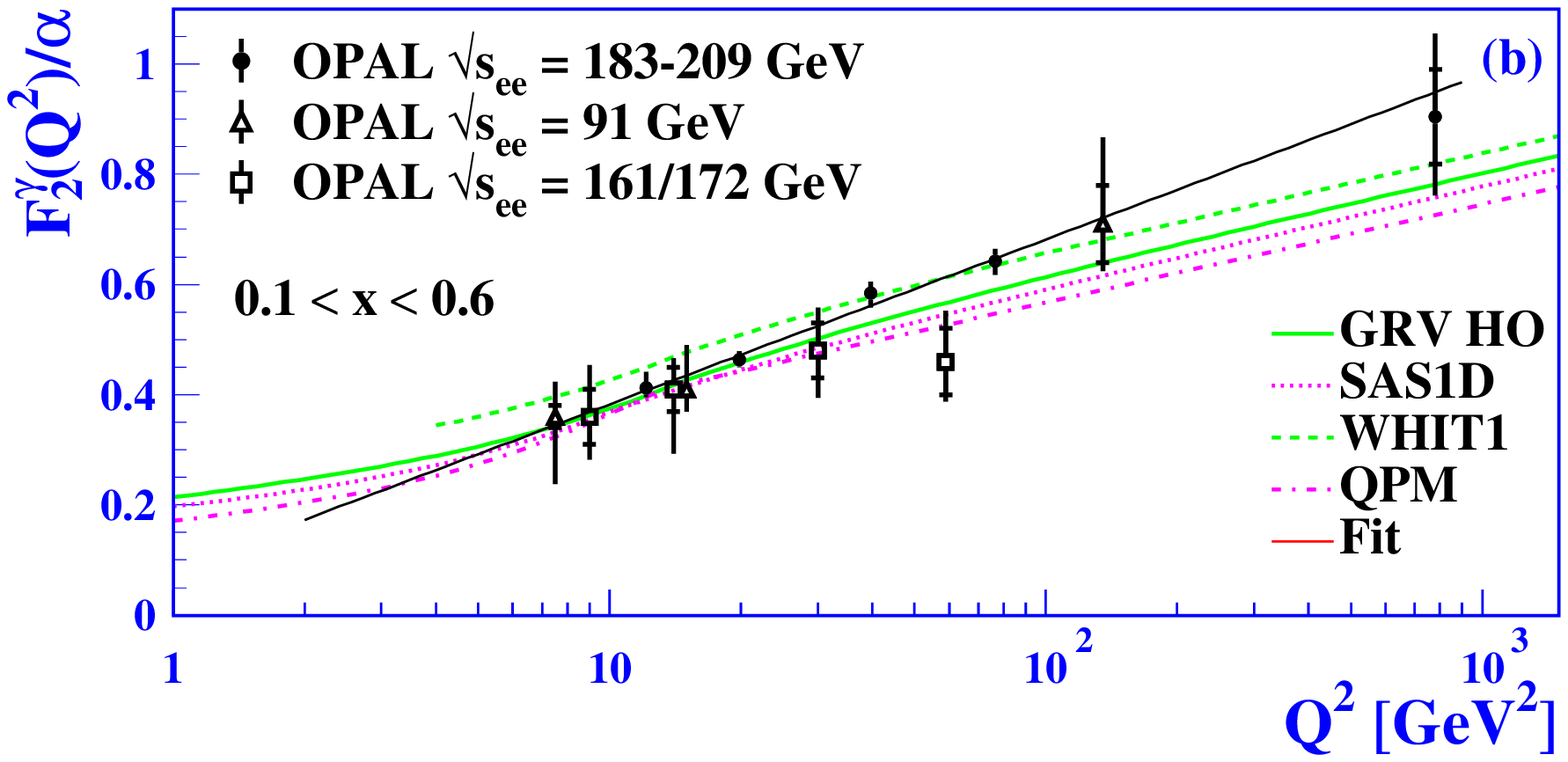}}
\caption{The evolution of \ftn as a function of \qsq for several bins
         of $x$, (a) 0.10--0.25, 0.25--0.60 and 0.60--0.85
         and (b) for the central region 0.10--0.60.
         The inner error bars represent the statistical errors and the outer 
         error bars represent statistical and systematic errors added in
         quadrature.
         In (a) the data are compared
         to the predictions from the GRSc (full line), SaS1D (dotted line),
         WHIT1 (dashed line), and QPM (dot-dashed line) 
         parameterisations of \ftn.
         In (b) GRSc has been replaced by the higher order prediction from
         GRV and, in addition, the result of the fit is shown.
        }\label{fig:fig05}
\end{center}
\end{figure}

 Following the study of the scaling violation of \ft performed 
 in~\cite{OPALPR207} the evolution of \ft with \qsq has been measured 
 for several $x$ ranges using all three samples.
 Due to their large statistics, the SW and FD samples are further split into 
 two bins of \qsq (9--15 and 15--30~\gevsq for SW
 and 30--50 and 50--150~\gevsq for FD).
 The data are unfolded as a function of $x$ separately in each bin of
 \qsq and corrected for radiative effects.
 The results are shown in Figure~\ref{fig:fig05} and listed in 
 Table~\ref{tab:resmea}.
 The estimation of the systematic errors for the EE sample is described above.
 For the SW and FD samples the estimation of the systematic error mirrors
 the procedure for the EE sample, with some differences.
 The PHOJET program is used as a second Monte Carlo to determine the 
 model dependence; the variations of cuts are given in Table~\ref{tab:cuts}.
 For the unfolding parameters the standard number of bins, which was 8 
 for the central values, has been varied by $\pm 2$.
 No systematics due to the electron isolation are needed for the SW
 and FD samples.
 For these samples, the largest contribution to the systematic 
 error generally stems from the estimated model dependence.
 With only two models available that satisfactorily describe the 
 data~\cite{OPALPR314}, the estimated systematic error is small for those
 $x,\qsq$ regions where the two models happen to predict similar correlations
 between $x$ and \xvis.
 To reduce fluctuations within the SW and FD samples, the systematic 
 error from this source has been averaged for each region of $x$ for the 
 two \qsq points within a given sample.
 \par
 The data in Figure~\ref{fig:fig05}(a) show positive scaling violations
 in \ft for the $x$ ranges 0.10--0.25 and 0.25--0.60.
 The QCD inspired parametrisations of \ft qualitatively follow the data,
 but do not perfectly account for them. 
 For the SW sample the GRSc and SaS1D predictions, which are almost 
 indistinguishable, closely resemble the data, whereas at higher \qsq, for 
 the FD sample, WHIT1 comes closest.
 For the range 0.60--0.85, the data are compatible 
 with the predicted scaling violations of the QCD inspired parametrisations.
 The QPM model generally gives a bad description of the data,
 especially at low $x$.
 \par
 To quantify the slope for medium values of $x$, where data are available
 at all values of \qsq, the data are fitted using essentially the 
 procedure from~\cite{OPALPR207}.
 A linear function of the form $a + b\,\ln\,\qsq$, where \qsq is in \gevsq,
 has been fitted to the data in the region 0.10--0.60.
 Within this range of $x$ the parameters $a$ and $b$ are assumed to be
 independent of $x$.
 To obtain the central values of the two parameters, with their statistical
 errors and correlation, a fit was performed by the MINUIT~\cite{JAM-9501} 
 program using the measured values of \ftn and their statistical errors 
 as listed in Table~\ref{tab:resmea}.
 The fit was repeated for each of the systematic variations.
 The systematic errors of $a$ and $b$ are estimated as the quadratic sum of
 the deviations of the two parameters from the central values.
 The result of the fit is $$\fitres\,,$$
 with, for the central result, a correlation between the two parameters 
 of $-0.98$ and a \chiq of 10 for 3 degrees of freedom.
 No significant change of the result is observed
 if the fit is performed using the full error on each point.
 This new result compares to the previous OPAL value~\cite{OPALPR207} of 
 $$\fitold.$$
 These two determinations, based on independent data sets, are in agreement,
 and the errors on $a$ and $b$ have been significantly reduced.
 The data, together with the fit result, are shown in 
 Figure~\ref{fig:fig05}(b).
 They are qualitatively described by the higher order GRV parametrisation
 (GRV HO).
%
%
\section{Conclusions}
\label{sec:concl}
 The photon structure function \ft and the differential cross-section \dsdx 
 have been measured using deep inelastic
 electron-photon scattering events recorded by the OPAL detector during the
 years 1997--2000 with an integrated luminosity of 632~\invpb and
 an average \epem centre-of-mass energy of 197.1~GeV.
 \par
 The structure function \ft has been measured as a function of $x$ in the
 range $0.15<x<0.98$ and at an average photon virtuality of 
 $\qzm=780$~\gevsq, which represents the highest value measured so far.
 The \qsq evolution of \ft has been studied for $12.1 < \qzm < 780$~\gevsq
 using several ranges of $x$.
 The data exhibit positive scaling violations in \ft for the $x$ ranges 
 0.10--0.25 and 0.25--0.60.
%
 For the range 0.60--0.85, the data are compatible with the predicted 
 scaling violations.
 The measured evolution of \ftn as a function of \qsq in the central region 
 of $x$, 0.10--0.60, has been fitted with a linear 
 function in $\ln\qsq$, resulting in $$\fitres\,,$$ where \qsq is in \gevsq.
 \par
 Both for the measurement of \ft at $\qzm=780$~\gevsq and for the 
 investigation of the \qsq evolution of \ft, the quark-parton model 
 prediction is not in agreement with the data.
 It shows a much steeper rise  than the data as a function of $x$ 
 for $\qzm=780$~\gevsq and also a different behaviour in the \qsq evolution.
 In contrast, the leading order GRSc, SaS1D and WHIT1 parameterisations and
 the higher order GRV parameterisation of \ft are much closer to the data.
 This means that the corresponding parton distribution functions of the 
 photon are adequate to within about 20$\%$ at large values of $x$ and at
 \qzm scales of about 780~\gevsq.
\clearpage
%
%
\appendix
\par
Acknowledgements:
\par
We particularly wish to thank the SL Division for the efficient operation
of the LEP accelerator at all energies
 and for their close cooperation with
our experimental group.  We thank our colleagues from CEA, DAPNIA/SPP,
CE-Saclay for their efforts over the years on the time-of-flight and trigger
systems which we continue to use.  In addition to the support staff at our own
institutions we are pleased to acknowledge the  \\
Department of Energy, USA, \\
National Science Foundation, USA, \\
Particle Physics and Astronomy Research Council, UK, \\
Natural Sciences and Engineering Research Council, Canada, \\
Israel Science Foundation, administered by the Israel
Academy of Science and Humanities, \\
Minerva Gesellschaft, \\
Benoziyo Center for High Energy Physics,\\
Japanese Ministry of Education, Science and Culture (the
Monbusho) and a grant under the Monbusho International
Science Research Program,\\
Japanese Society for the Promotion of Science (JSPS),\\
German Israeli Bi-national Science Foundation (GIF), \\
Bundesministerium f\"ur Bildung und Forschung, Germany, \\
National Research Council of Canada, \\
Research Corporation, USA,\\
Hungarian Foundation for Scientific Research, OTKA T-029328, 
T023793 and OTKA F-023259,\\
Fund for Scientific Research, Flanders, F.W.O.-Vlaanderen, Belgium.\\
%
%

%
\clearpage
%
%
\renewcommand{\arraystretch}{1.10}
\begin{table}[htb]
\begin{center}\begin{tabular}{|c|c|c|c|}\hline
  Cut $\backslash$ Sample           &      SW &             FD  &   EE \\
 \hline
 \etag/\eb  min        & \multicolumn{2}{c|}{$0.75\,(\pm 0.05)$}
                       & $0.70\,(\pm 0.05)$ \\
 \hline
 \ttag      min [mrad] & $33.25\,(+2)$ &  $60\,(+3)$ & $230\,(\pm 5)$  \\
 \hline
 \ttag      max [mrad] & $55\,(-2)$    & $120\,(-3)$ & $500\,(\pm 5)$  \\
 \hline
 \ea/\eb    max        & \multicolumn{3}{c|}{$0.25\,(\pm 0.05)$}        \\
 \hline
 \nch       min        & \multicolumn{2}{c|}{$3\,(+1)$}    & $4\,(+1)$  \\ 
                       & \multicolumn{3}{c|}{(2 non-electron tracks)}\\
 \hline
 \Wvis    min $[\gev]$ & \multicolumn{3}{c|}{$2.5\,(+1)$}         \\
 \hline
 \Wvis    max $[\gev]$ & \multicolumn{2}{c|}{$60\,(\pm 5)$} & $50\,(\pm 5)$  \\
 \hline
 Electron isolation $[\gev]$& \multicolumn{2}{c|}{-}      & $3.0\,(\pm 0.5)$\\
 \hline
 \end{tabular}
 \caption{The selection cuts applied to each data sample, together with the 
          variations applied (in brackets).
          See the text for explanation of the variables.
         }\label{tab:cuts}
\end{center}\end{table}
%
\renewcommand{\arraystretch}{1.10}
\begin{table}[htb]
\begin{center}\begin{tabular}{|c|c|r@{$\,\pm\,$}l|r@{$\,\pm\,$}l|r@{$\,\pm\,$}l|}\hline
\multicolumn{2}{|c|}{} & \multicolumn{2}{|c|}{SW} & \multicolumn{2}{|c|}{FD} & \multicolumn{2}{|c|}{EE}\\\hline
\multicolumn{2}{|c|}{data selected}   & \multicolumn{2}{|c|}{27819} & \multicolumn{2}{|c|}{11874} & \multicolumn{2}{|c|}{414}         \\\hline
\multicolumn{2}{|c|}{data signal} & 26071&167 & 10652&110 & 274&21 \\\hline
\multicolumn{2}{|c|}{\hspace{3mm} Monte Carlo selected} & 
             28308&51 & 11211&32 & 436&6\\\hline
\multicolumn{2}{|c|}{\hspace{3mm}HERWIG signal} & 
               26560&49  &  9989&30  & 296&5  \\\hline
 & \ggtau   & 1309.3&14.1 & 845.5&11.3 & 31.8&2.2  \\\cline{2-8}
 & \gsgshad &  321.3&4.7  & 193.4&3.7  &  5.0&0.3  \\\cline{2-8}
 & \znhad &   82.8&2.4  & 124.6&3.1  & 76.2&2.4  \\\cline{2-8}
 & \zntau &    7.9&0.3  &  10.5&0.4  & 10.6&0.4  \\\cline{2-8}
\hspace{2.3mm}\begin{rotate}{90}$\,$Backgrounds\end{rotate} 
 & 4-fermion eeqq &   27.0&0.9  &  48.2&1.1  & 16.6&0.7  \\\hline
\end{tabular}
\caption{The numbers of selected events and signal events (selected events
         corrected for background) in the data compared to the signal 
         predictions from the HERWIG program. 
         The expected numbers of background events for the dominant sources
         according to Monte Carlo are also listed. 
         The errors given are only statistical.
        }\label{tab:samples}
\end{center}\end{table}
%
\renewcommand{\arraystretch}{1.10}
\begin{table}[t]
\begin{center}
\begin{tabular}{|c|c|c|c|c|c|}\hline
    $x$ & $x$        & \ftn & \dsdx & radiative         & bin-centre \\
  range & bin-centre &      & [pb]  & cor. [$\%$] & cor. [$\%$] \\\hline
$0.15-0.40$ & $0.275$ &\Z{0.93}{0.10}{0.14}{0.11}&\Z{0.94}{0.11}{0.15}{0.11}& $-8.8$ & $-0.4$ \\
$0.40-0.70$ & $0.550$ &\Z{0.87}{0.10}{0.05}{0.15}&\Z{0.79}{0.09}{0.05}{0.14}& $-6.9$ & $\pz0.1$\\
$0.70-0.98$ & $0.840$ &\Z{0.97}{0.17}{0.16}{0.23}&\Z{0.62}{0.11}{0.10}{0.15}& $-5.4$ & $-5.1$ \\
\hline 
\end{tabular}
\mbox{ }\\\vspace{1cm}
\begin{tabular}{|c|c|c|c|}\hline
 $x$ range   & $0.15-0.40$   & $0.40-0.70$ & $0.70-0.98$ \\\hline
 $0.15-0.40$ &  $\pz 1.00$   &             &             \\\hline
 $0.40-0.70$ &  $\pz 0.16$   & $\pz 1.00$  &             \\\hline
 $0.70-0.98$ &  $   -0.04$   & $   -0.15$  & $1.00$      \\\hline
\end{tabular}
\caption{Results for the EE sample for \ftn as a function of $x$ at \qzm of 
         780~\gevsq, and for \dsdx in the \qsq range 400-2350~\gevsq and
         $y<0.3$. 
         The first errors are statistical and the second systematic.
         The data were unfolded in bins defined by the
         $x$ ranges and corrected for radiative effects.
         For a given bin the radiative correction is the difference of the
         radiative and non-radiative cross-sections as a percentage
         of the non-radiative cross-section.
         The structure function was corrected to the $x$ values listed
         using the bin-centre corrections which are given as a percentage
         of the non-corrected \ft.
         The statistical correlations between the bins for
         the central result are also given.
        }\label{tab:results}
\end{center}\end{table}
%
\renewcommand{\arraystretch}{1.20}
\begin{table}[t]
\begin{center}
\begin{tabular}{|c|c|c|c|c@{$\,\pm\,$}c@{$\,^{+\,}_{-\,}$}c|c|}
\multicolumn{6}{l}{(a)}\\
\hline
 $x$ range & \qsq range & \qzm     & \ftn & \multicolumn{3}{|c|}{\dsdx} & radiative  \\
           & $[\gevsq]$ & $[\gevsq]$ &    & \multicolumn{3}{|c|}{[pb]}  & cor. [$\%$]\\
\hline
 $0.10-0.25$ & $\po\po9-15\po\po$ & 12.1 &\Z{0.38}{0.01}{0.03}{0.03}& 83&1&\ZZ{7}{6} & $-5.3$ \\
 $0.25-0.60$ &                    &        &\Z{0.43}{0.01}{0.03}{0.02}& 47&1&\ZZ{4}{2} & $-4.4$ \\
\hline
 $0.10-0.25$ & $\po15-30\po\po$ & 19.9 &\Z{0.39}{0.01}{0.03}{0.03}& 56&1&\ZZ{4}{4} & $-5.5$ \\
 $0.25-0.60$ &                  &        &\Z{0.49}{0.01}{0.02}{0.02}& 36&1&\ZZ{1}{1} & $-4.5$ \\
\hline
 $0.10-0.25$ & $\po30-50\po\po$ & 39.7 &\Z{0.47}{0.01}{0.02}{0.02}& 21.7&0.4&\ZZ{0.9}{0.9} & $-5.9$ \\
 $0.25-0.60$ &                  &        &\Z{0.63}{0.01}{0.02}{0.03}& 15.9&0.4&\ZZ{0.5}{0.7} & $-4.8$ \\
 $0.60-0.85$ &                  &        &\Z{0.65}{0.03}{0.06}{0.06}&$\pw9.6$&0.4&\ZZ{0.9}{0.8} & $-3.8$ \\
\hline
 $0.10-0.25$ & $\po50-150\po$ & 76.4 &\Z{0.55}{0.01}{0.02}{0.03}& 18.8&0.4&\ZZ{0.7}{0.9} & $-6.5$ \\
 $0.25-0.60$ &                &        &\Z{0.68}{0.01}{0.02}{0.02}& 13.8&0.3&\ZZ{0.4}{0.4} & $-5.2$ \\
 $0.60-0.85$ &                &        &\Z{0.73}{0.02}{0.04}{0.04}&$\pw9.1$&0.3&\ZZ{0.4}{0.5} & $-4.1$ \\
\hline
 $0.25-0.60$ & $400-2350$ & 780  &\Z{0.94}{0.09}{0.09}{0.11}& 0.91&0.09&\ZZ{0.08}{0.10} & $-7.6$ \\
 $0.60-0.85$ &            &        &\Z{0.83}{0.11}{0.21}{0.29}& 0.71&0.09&\ZZ{0.18}{0.25} & $-6.0$ \\
\hline
\end{tabular}
\mbox{ }\\\vspace{3mm}
\begin{tabular}{|c|c|c|c|}\hline
 $x$ range   & $0.10-0.25$            & $0.25-0.60$    & $0.60-0.85$ \\\hline
 $0.10-0.25$ & 1                      &                      &   \\\hline
 $0.25-0.60$ & 0.00/0.28/0.45/0.40/-- & 1                    &   \\\hline
 $0.60-0.85$ & --/--/-0.23/-0.19/--   & --/--/0.32/0.27/0.17 & 1 \\\hline
\end{tabular}
\mbox{ }\\\vspace{3mm}
\begin{tabular}{|c|c|c|c|c@{$\,\pm\,$}c@{$\,^{+\,}_{-\,}$}c|c|}
\multicolumn{6}{l}{(b)}\\
\hline
 $x$ range & \qsq range & \qzm       & \ftn & \multicolumn{3}{|c|}{\dsdx} & radiative  \\
           & $[\gevsq]$ & $[\gevsq]$ &      & \multicolumn{3}{|c|}{[pb]}  & cor. [$\%$]\\
\hline
 $0.10-0.60$ & $\po\po9-15\po\po$ & 12.1 &\Z{0.41}{0.01}{0.03}{0.02}& 57&1&\ZZ{4}{2}      & $-4.7$ \\\po
             & $\po15-30\po\po$   & 19.9 &\Z{0.46}{0.01}{0.01}{0.01}& 42&1&\ZZ{1}{1}      & $-4.9$ \\
             & $\po30-50\po\po$   & 39.7 &\Z{0.58}{0.01}{0.02}{0.02}& 17.7&0.3&\ZZ{0.6}{0.7}    & $-5.3$ \\
             & $\po50-150\po$     & 76.4 &\Z{0.64}{0.01}{0.02}{0.02}& 15.3&0.3&\ZZ{0.4}{0.5}    & $-5.6$ \\
             & $400-2350$         & 780  &\Z{0.90}{0.09}{0.13}{0.11}& 0.86&0.08&\ZZ{0.12}{0.11} & $-8.2$ \\
\hline 
\end{tabular}
\caption{Results for the average \ftn in bins of $x$ for several values of 
         \qzm and for \dsdx for several ranges of \qsq.
         Section (a) corresponds to Figure~\protect\ref{fig:fig05}(a), 
         section (b) to Figure~\ref{fig:fig05}(b)
         The first errors are statistical and the second systematic.
         The data were unfolded in bins defined by the
         $x$ and \qsq ranges listed.
         The radiative corrections which have been applied are also listed.
         For a given bin the radiative correction is the difference of the
         radiative and non-radiative cross-sections as a percentage
         of the non-radiative cross-section.
         The statistical correlations between the bins for
         the central result for the \qzm values 12.1/19.9/39.7/76.4/780 \gevsq
         are also given.
        }\label{tab:resmea}
\end{center}\end{table}
%
 \clearpage
%
%
%
%
%
%
%
\end{document}